\documentclass[%
 reprint,
 amsmath,amssymb,
 aps,
]{revtex4-2}

\usepackage{braket}
\usepackage[caption=false]{subfig}
\usepackage{graphicx}
\usepackage{dcolumn}
\usepackage{physics}
\usepackage{bm}%

\renewcommand{\epsilon}{\varepsilon}

\renewcommand{\l}{\left}
\renewcommand{\r}{\right}

\newcommand{\tx}{\text}

\newcommand{\nn}{\nonumber}

\newcommand{\rom}[1]{\uppercase\expandafter{\romannumeral #1\relax}}

\def\beq{\begin{equation}}
\def\endeq{\end{equation}}

\def\ba#1\ea{\begin{align}#1\end{align}}

\def\bs{\begin{subequations}}
\def\es{\end{subequations}}

\begin{document}

\title{Directed percolation in non-unitary quantum cellular automata}

\author{Ramil Nigmatullin}
\email{ramil.nigmatullin@mq.edu.au}
\author{Elisabeth Wagner}%
\author{Gavin K. Brennen}
 \affiliation{Center for Engineered Quantum Systems, Dept.~of Physics \& Astronomy, Macquarie University, 2109 NSW, Australia}

\date{\today}

\begin{abstract}
Probabilistic cellular automata (CA) provides a classic framework for studying non-equilibrium statistical physics on a lattices.
A notable example is the Domany-Kinzel CA, which has been used to investigate the process of directed percolation and the critical dynamics of the non-equilibrium phase transition betweeen absorbing and percolating phases. In this work, we construct a non-unitary Quantum Cellular Automaton that generalises the Domany-Kinzel cellular automaton and study the resulting dynamical evolution using the numerical simulations using the tensor network iTEBD algorithm. We demonstrate the system undergoes the absorbing/percolating phase transition and the addition of the Hamiltonian generates coherences, which are a distinct feature of the quantum dynamics. A proposal for the implementation of the model with Rydberg array is put forward, which does not require local addressing of individual sites. 
\end{abstract}

\maketitle

\section{Introduction}

In recent years there have been great advances in the development of quantum simulation platforms. These include ultracold atoms, ions, superconducting qubits and photonic systems. One of the most recent advances has been the development of a quantum simulators based on arrays of ultracold Rydberg atoms \cite{Browaeys2020}. Atoms that are excited to Rydberg states interact strongly, which makes it possible to engineer controlled multiqubit dynamics. Rydberg arrays provide excellent platform for studying non-equilibrium dynamics of quantum many-body systems, for example, for probing non-equilibrium quench dynamics in Ising spin chains \cite{Bernien2017,PhysRevX.8.021069}, energy transport \cite{PhysRevLett.114.113002,deLeseleuc775} and the validity of thermalization hypothesis \cite{PhysRevLett.120.180502,Turner2018}. 

In the domain of non-equilibrium physics, one promising application of Rydberg arrays is in exploration of non-equilibrium phase transitions (NEPTs), the study of which is particularly challenging due to the requirement of large system sizes and long time evolution needed to reach steady states. The classical NEPTs are classified into universality classes characterized by power scaling laws at critical points \cite{doi:10.1080/00018730050198152,RevModPhys.76.663}. 
Quantum fluctuations can drastically alter the nature of NEPT, for example, it was shown that the critical exponents of NEPT in a quantum analogue of the contact process model are different to the critical exponents of a classical model \cite{Gillman_2019}. The quantum generalization of NEPT theory remains largely unexplored both theoretically and experimentally, because of the challenges in simulating the dynamics of quantum many-body system and engineering such systems in the lab. 

Perhaps the most fundamental NEPT is the phase transition between absorbing and percolating phase in lattice systems undergoing directed percolation (DP) dynamics \cite{doi:10.1080/00018730050198152}. A model generalizing DP into the quantum domain, which can be implemented using 2d Rydberg arrays, has been recently proposed \cite{doi:10.1080/00018730050198152} and studied numerically \cite{PhysRevLett.125.100403,Gillman2020}. The model in \cite{doi:10.1080/00018730050198152} involves application of local 3-site unitary gates, which sequentially updates the rows, such that the state of each row depends on the state of the previous row and corresponds to the state of a probabilistic cellular automaton undergoing DP dynamics. 

In this paper, we construct a quantum generalization of DP, which uses a combination of non-trivial dissipative dynamics as well as coherent nearest neighbor dynamics. The scheme is inspired by the recently developed techniques for engineering dissipative interactions in the Rydberg arrays \cite{PhysRevLett.124.070503}. The strength of the influence of quantum effects can be varied by tuning the relative contribution of coherent and stochastic terms to the overall dynamics. The proposed scheme has several advantages. In particular, it does not require addressing of the individual lattice sites as all of the sites can be updated simultaneously with spatially homogeneous dynamics. 
The scheme can also be viewed as a robust way of preparing many-body quantum states, since the reached steady states are independent of the initially prepared state. The approach can be easily generalized to (2+1)-dimensional quantum cellular automata for probing more complex non-equilibrium models. 

The paper is organized as follows. Section II describes the construction of the non-unitary QCA rules that generate the DP process. Section III describes the simulation method and section IV presents the results and draw the comparison with the classical stochastic DP. Finally, the possible physical implementation in Rydberg arrays is discussed in section V.

\section{Model}

We consider a 1-dimensional spin lattice, which undergoes open quantum dynamical evolution modeled using a Markovian master equation. Both the Hamiltonian and the Lindblad jump operator in the master equation are 3-qubit operators coupling the nearest neighboring sites in the array. The master equation is given by

\begin{equation}
    \partial_t \rho = \mathcal{L}[\rho] = - i [H, \rho] +\mathcal{D}[\rho], \label{eq:master}
\end{equation}
where the Hamiltonian is of the form

\begin{equation}
    H = \frac{1}{2} \sum_j \sum_{\alpha\beta} \theta_{\alpha \beta} \dyad{\alpha}_{j-1} \otimes X_j \otimes \dyad{\beta}_{j+1}, \label{eq:Hmodel}
\end{equation}
where $\alpha,\beta\in\{0,1\}$, $\dyad{\alpha}_j$ is a projector on site $j$, $X_j$ is the Pauli-X operator on site $j$ and $\theta_{\alpha\beta}$ are constants whose values can be experimentally tuned. 

The dissipator is given by

\begin{eqnarray}
    \mathcal{D}[\rho] = &\sum_j \left[L^+_j \rho L^{+\dagger}_j - \frac{1}{2} \left(L^{+\dagger}_j L^+_j \rho +\rho L^{+\dagger}_j L^+_j \right) \right. \nonumber\\
      & \left.+ L^-_j \rho L^{-\dagger}_j - \frac{1}{2} \left(L^{-\dagger}_j L^-_j \rho +\rho L^{-\dagger}_j L^-_j \right)\right]
\end{eqnarray}
with the jump operators

\begin{equation}
    L_j^{\pm} = \sum_{\alpha\beta} \sqrt{\gamma_{\alpha\beta}^\pm}\dyad{\alpha}_{j-1} \otimes \sigma^{\pm}_j\otimes\dyad{\beta}_{j+1}, \label{eq:Lmodel}
\end{equation}
where $\sigma^\pm_j=X_j\pm i Y_j$ is a lowering/raising operators on site $j$, $\gamma_{\alpha \beta}^\pm$ are tunable rate constants. If the Hamiltonian term is set to zero, $\theta_{\alpha\beta}=0$ for all $\alpha$, $\beta\in\{0,1\}$, then the master equation describes classical stochastic dynamics on a lattice since no coherence can be generated if the system is initialized in a product state. Such dynamics can be realized in atomic systems such as trapped ions and Rydberg atom arrays. A possible implementation in a Rydberg array will be described in section \ref{sec:implementation}. 
\begin{figure}[t]
    \centering
    \includegraphics[scale=0.75]{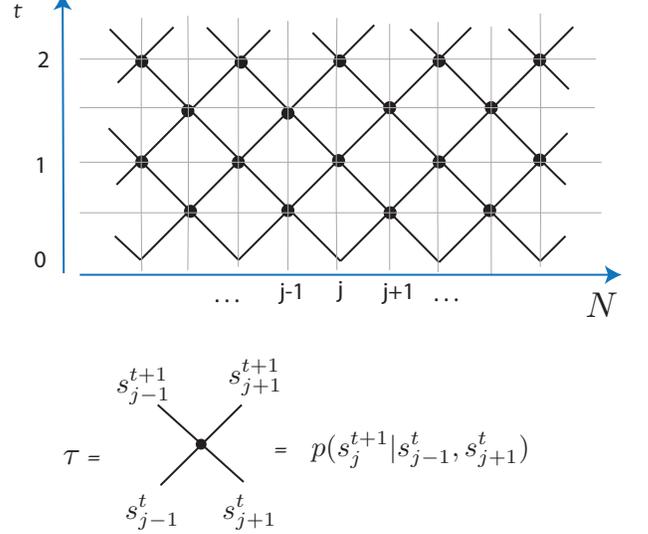}
    \caption{A schematic representation of (1+1)-dimensional block partitioned CA with nearest neighbor update rule, e.g. Domany-Kinzel CA. The state of a system at time $t$ is given by contracting all the transfer matrix tensors $\tau$ in the preceding time steps.}
    \label{fig:DKCA}
\end{figure}

We will now address the question of how to appropriately choose the rates $\gamma_{\alpha \beta}^\pm$, such that the model would correspond to the process of directed percolation (DP) in the fully stochastic incoherent regime. The idea is to choose the Lindblad operators such that the generated dynamics can be identified with the evolution of the stochastic Domany-Kinzel cellular automaton (DKCA) \cite{PhysRevLett.53.311}. DKCA is a model exhibiting the NEPT in the DP universality class, which is particularly simple to simulate computationally and has been used extensively to study the absorbing/percolating critical point. DKCA is a (1+1)-dimensional discrete stochastic CA, where the state of each cell is updated with a probability that depends on the state of its left and right neighbors. The probabilistic update rule $p(s_j^t|s_{j-1}^t,s_{j+1}^t)$ is given by $p(1|0,0)=x$, $p(1|0,1)=p(1,1,0)=y$, $p(1|1,1)=z$ and $p(0|s_{j-1}^t,s_{j+1}^t)=1-p(1|s_{j-1}^t,s_{j+1}^t)$. DKCA is block partitioned such that the updates of its even and odd cells is alternated, as shown in figure \ref{fig:DKCA}. When $x=0$, the dynamics of the DKCA corresponds to DP. It is not hard to see that setting $y=z=p$ maps the dynamics to (1+1)-dimensional site percolation processes, where $p$ is the probability presence of a site. The model corresponds to bond percolation, when setting $y=q$ and $z=q(2-q)$, where $q$ is the probability of presence of a bond.

For our model to correspond to DKCA stochastic rules, we will require that the site occupations in the stationary state of the CP map acting on an isolated 3-cell (a site and its neighbors) are the same as in the stationary states of the DKCA rule. Let $T$ be the transfer matrix of a DKCA 3-cell rule. Let $\textbf{v}$ be the eigenvector of $T$ with the identity eigenvalue, $T\textbf{v}=\textbf{v}$. The vector $\textbf{v}$ is the stationary state of the DKCA 3-cell rule. The stationary state of the CP map generated by the master equation satisfies $e^{-\mathcal{L}t}[\rho^s]=\rho^s$ for all $t$; or, equivalently, $\mathcal{L}[\rho^s]=0$.
We choose parameters $\{\gamma_{\alpha\beta}^\pm, \theta_{\alpha\beta} \}$ such that the diagonal elements of $\rho^s$ are equal to the elements $\textbf{v}$, i.e.~$\textrm{diag}(\rho^s)=\textbf{v}$. The rates $\gamma^\pm_{\alpha\beta}$ are then adjusted according to
\begin{widetext}
\begin{eqnarray}
\gamma^+_{\alpha\beta} =
\begin{cases}
 \frac{p_{\alpha\beta}}{1-p_{\alpha\beta}}\,\gamma_{\alpha\beta}^- &\text{if }\theta_{\alpha\beta} = 0 
 \\
\frac{(1-2p_{\alpha\beta})\gamma^-_{\alpha\beta}+ \sqrt{(\gamma^-_{\alpha\beta})^2-16 p_{\alpha\beta} \theta^2_{\alpha\beta}(1-3p_{\alpha\beta}+2p_{\alpha\beta}^2)}}{2(1-p_{\alpha\beta})} &\text{if }\theta_{\alpha\beta}\neq 0
\end{cases},\label{eq:gammap}
\end{eqnarray}
\end{widetext}
whose derivation is presented in appendix \ref{sec:app}.

\pagebreak
Note that for the model to be physical, the jump rates have to be positive and real. This puts a restriction on the range of allowed values of $p_{\alpha\beta}$ and $\theta_{\alpha\beta}$. If we set $\gamma_{\alpha\beta}^-=1$, then we find that $\gamma_{\alpha\beta}^+$ is real and positive for
$p_{\alpha\beta}\geq\frac{1}{2} \ \forall\,\theta_{\alpha\beta}$, or for 
$p_{\alpha\beta}<\frac{1}{2}$ if 
$\theta_{\alpha\beta} \leq \frac{1}{16(1-3p_{\alpha\beta}+2p_{\alpha\beta}^2)}$.


\begin{figure}
    \centering
    \includegraphics[scale=0.5]{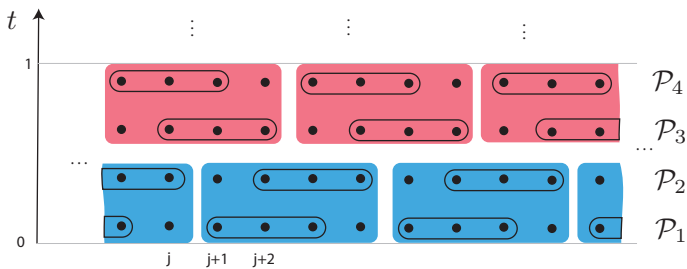}
    \caption{Block partitioning scheme of the proposed non-unitary QCA model of directed percolation. Partitions $\mathcal{P}_1$ and $\mathcal{P}_4$ update even cells and partitions $\mathcal{P}_2$ and $\mathcal{P}_3$ update odd cells.}
    \label{fig:partitioning}
\end{figure}

Having constructed the master equation whose dynamics implements the quantum analogue of the DKCA 3-cell update rule, the remaining task is to define the order in which the updates are to be carried out. We propose to split the updates into four partitions, $\mathcal{P}_1$, $\mathcal{P}_2$, $\mathcal{P}_3$ and $\mathcal{P}_4$ as shown in figure \ref{fig:partitioning}. All the cells in each partition can be updated in parallel, since the 3-cell CP maps with each partition act on distinct cells with no overlaps.

\section{Numerical method and calculation setup}

The simulation method used in this paper is based on representing the state of the system as matrix product state (MPS) implementing time evolution using infinite time-evolving block decimation (iTEBD). MPS methods have been demonstrated to be efficient tools for studying one dimensional quantum systems \cite{RevModPhys.77.259}. If a system has translational invariance then the iTEBD method can be used to compute the dynamics of effectively infinite system. Though iTEBD has mostly been used to compute the unitary dynamics of closed quantum system, it can also be applied to open dynamics with dissipation and decoherence \cite{PhysRevB.78.155117}. 

To  apply iTEBD to Lindblad dynamics given by equation (\ref{eq:master}), we will vectorize the density matrix using the Choi-isomorphism $\dyad{\alpha}{\beta}\rightarrow \ket{\alpha}\otimes\ket{\beta}$. Mapped in this way the density matrix $\rho(t)=\sum_{ij} \rho_{ij}(t)\dyad{i}{j}$ becomes a vector in a doubled-space $\ket{\rho(t)}=\sum_{ij} \rho_{ij}(t) \ket{i}\otimes\ket{j}$. Under this mapping the Lindblad equation (\ref{eq:master}) becomes
\begin{equation}
    \frac{d}{dt} \ket{\rho(t)} = \mathbb{L} \ket{\rho(t)},\label{eq:mastervec}
\end{equation}
where the Lindblad map $\mathbb{L}$ in doubled space is given by
\begin{equation}
    \mathbb{L} = \mathbb H+\mathbb{D}, \label{eq:vectorized}
\end{equation}
with 
\begin{equation}
    \mathbb H = -i(H\otimes I- I\otimes H^T),\label{eq:Hvec}
\end{equation}
and 
\begin{widetext}
\begin{equation}
    \mathbb{D} = \sum_\pm\sum_j \left( L_j^\pm \otimes (L_j^\pm)^* - \frac{1}{2}\left[(L_j^\pm)^\dagger L_j^\pm \otimes I + I\otimes (L_j^\pm)^T (L_j^\pm)^* \right] \right).\label{eq:Dvec}
\end{equation}
\end{widetext}

Solution to equation (\ref{eq:master}) is given by $\ket{\rho(t)}=e^{\mathbb{L} t} \ket{\rho(0)}$, where $\ket{\rho(0)}$ is the initial state in the vectorized representation. 
Defining $\mathbb{L}_j$ as the Liouvillian acting on the cells in the partition $\mathcal{P}_j$, the full round of the discrete non-unitary QCA model updating the state at step $s$ to step $s+1$ is given by
\begin{equation}
    \ket{\rho_{s+1}} = e^{\mathbb{L}_4 \tau}e^{\mathbb{L}_3 \tau}e^{\mathbb{L}_2 \tau}e^{\mathbb{L}_1 \tau} \ket{\rho_s},
\end{equation}
where the time $\tau$ is the time for which the system continuously evolves before switching to the next partition. If $\tau$ is large relative to the characteristic time of the dissipation, i.e. $\tau \gg 1/\sqrt{\gamma}$, then the three cells undergoing the dynamical evolution reach a steady state, which corresponds to the state of the DKCA in the purely stochastic regime. 
If $\tau$ is small, i.e.~$\tau\rightarrow 0$, then the partitioning can be viewed as 
Trotter decomposition of the time evolution $e^{\mathbb{L} t}$, where $\mathbb{L}$ acts on all sites simultaneously. Thus, in the limit $\tau\rightarrow 0$ the model can be regarded as a continuous non-unitary QCA, as the evolution happens continuously rather than in discrete steps while the notion of locality is still being preserved, as both the Hamiltonian and the Lindblad jump operators are supported over nearest neighbors. 

\begin{figure}
    \centering
    \subfloat[]{
    \includegraphics[scale=0.4]{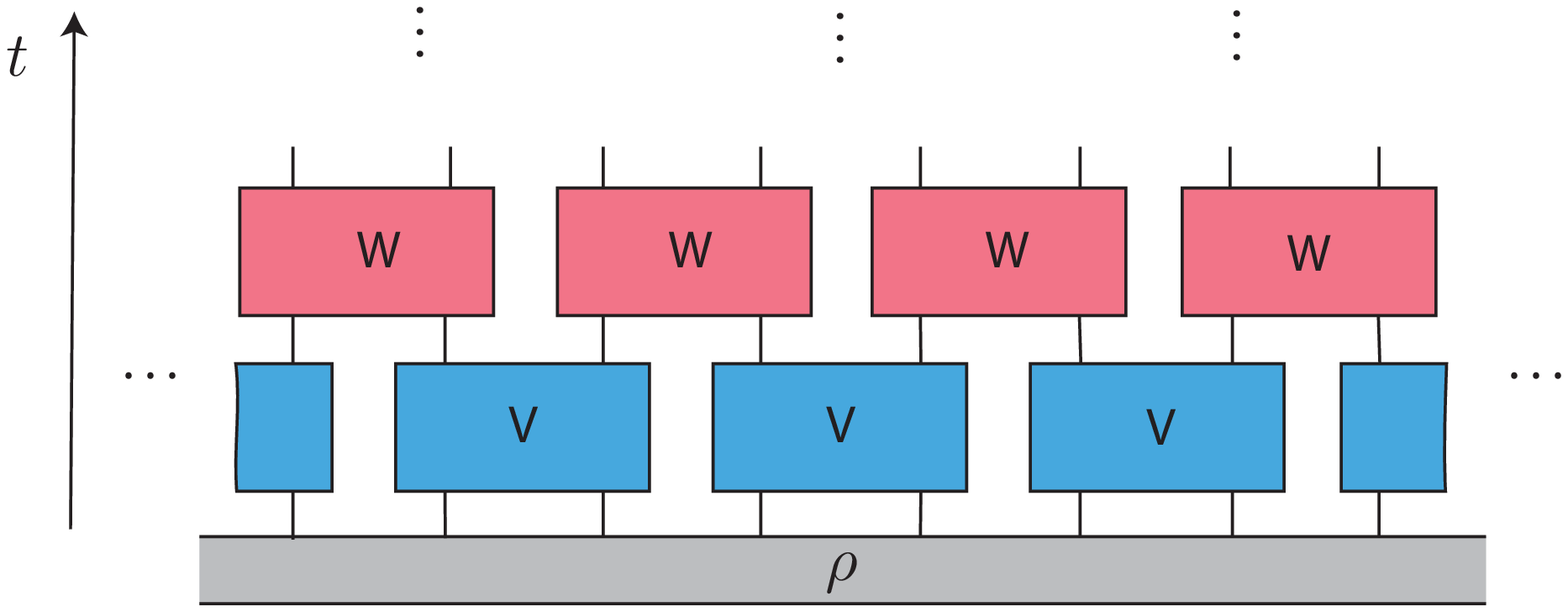}}
    
    \subfloat[]{
    \includegraphics[scale=0.45]{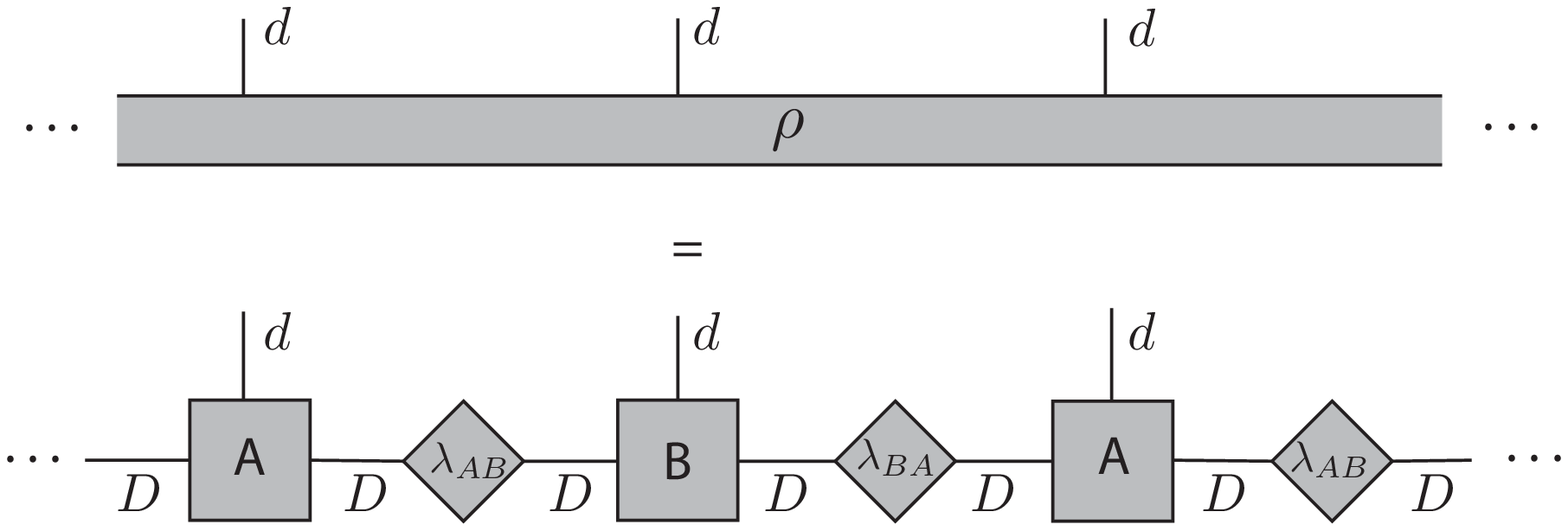}}
    \caption{(a)  Tensor network representation of the single round of the QCA update with tensors $V$ and $W$ given by $V\equiv e^{\mathbb{L}_2 \tau} e^{\mathbb{L}_1 \tau}$ and $W=e^{ \mathbb{L}_4 \tau} e^{\mathbb{L}_3 \tau}$. (b) MPS representation of the state vector $\ket{\rho}$. The physical index dimension is $d=16$ and the bond dimension $D$ is chosen to obtain the desired accuracy and speed of the iTEBD algorithm.}
       \label{fig:TN}
\end{figure}

Figure \ref{fig:TN} shows the tensor network representation of the time evolution, where the operator $V\equiv e^{\mathbb{L}_2 \tau} e^{\mathbb{L}_1 \tau}$ and $W=e^{\mathbb{L}_4 \tau} e^{\mathbb{L}_3 \tau}$. The state $\rho$ is represented as an MPS with dimension $D$. 
Application of the $V$ and $W$ completely positive maps increases the bond dimension of $\rho$. In TEBD algorithms the growth of the bond dimension is controlled by only keeping a specified number of Schmidt values.  From the approximate $\ket{\rho(t)}$, one can then compute the observables $\hat{O}$ 
\begin{equation}
    O(t) = \langle \hat{O}\rangle = \mathrm{tr} \left[ \rho(t)\hat{O}_D \right] = \langle \mathbb{I}|\hat{O}_D|\rho(t)\rangle,
\end{equation}
where $\hat{O}_D = \hat{O}\otimes I$ and $\ket{\mathbb{I}}$ is the double space representation of the identity operator. 

\section{Simulations results and discussion}

\subsection{Purely stochastic limit}
To verify that our non-unitary QCA model indeed represents the processes of directed percolation on a lattice, we first compute the phase diagram of the model in the purely stochastic regime, where $\theta_{\alpha\beta}=0$ for all $\alpha$, $\beta$. The initial state is set to  $\rho=\dyad{1}^{\otimes^\infty}$ i.e.~fully active state. The time evolution is computed using the iTEBD, for various decay rates $\gamma^+$ and $\gamma^-$. We set $\gamma^+=\gamma^-=p$ and vary the parameter $p$, which would correspond to the site percolation process. Both continuous and discrete QCA model is considered by simulating the dynamics with a 3-cell rule generated by Liouvillian evolution with short and long durations.
The lowest order Trotter error is proportional to the commutators of the jump operators $L^\pm$ and hence to keep this error constant for the continuous dynamic simulations across a range of $p$ values, we adjust the time steps such that $\tau^2 p (1-p)=C$ with the constant $C$ set to 0.0025. 
For discrete dynamics simulations, we set $\tau=10.0$, which is sufficient time for  a 3-cell to reach a steady state before a switch to the subsequent partitioning. 

\begin{figure*}
     \centering
     \subfloat[][]{
        \includegraphics[scale=0.33]{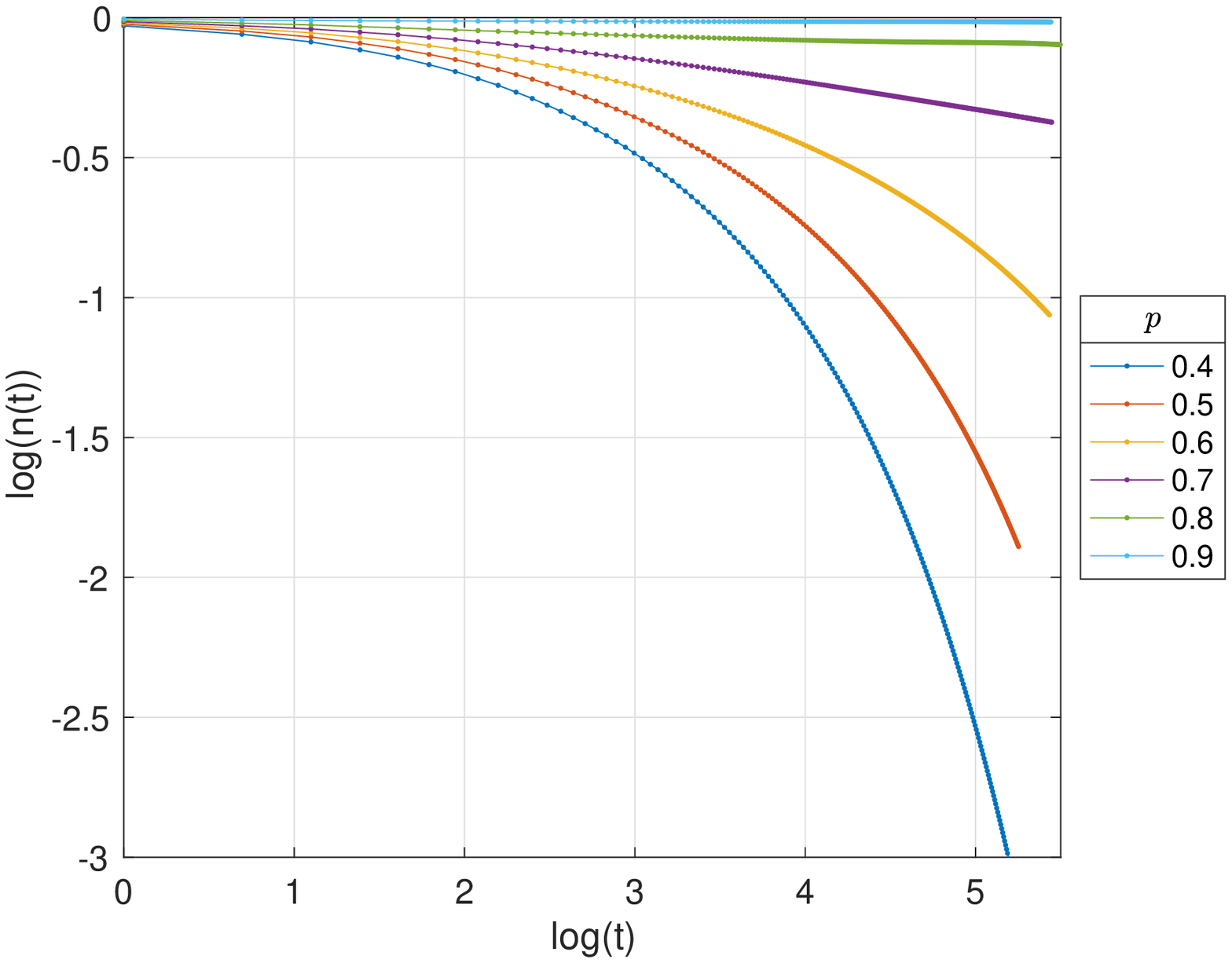}
     \label{fig1cont}}
    \subfloat[][]{
        \includegraphics[scale=0.33]{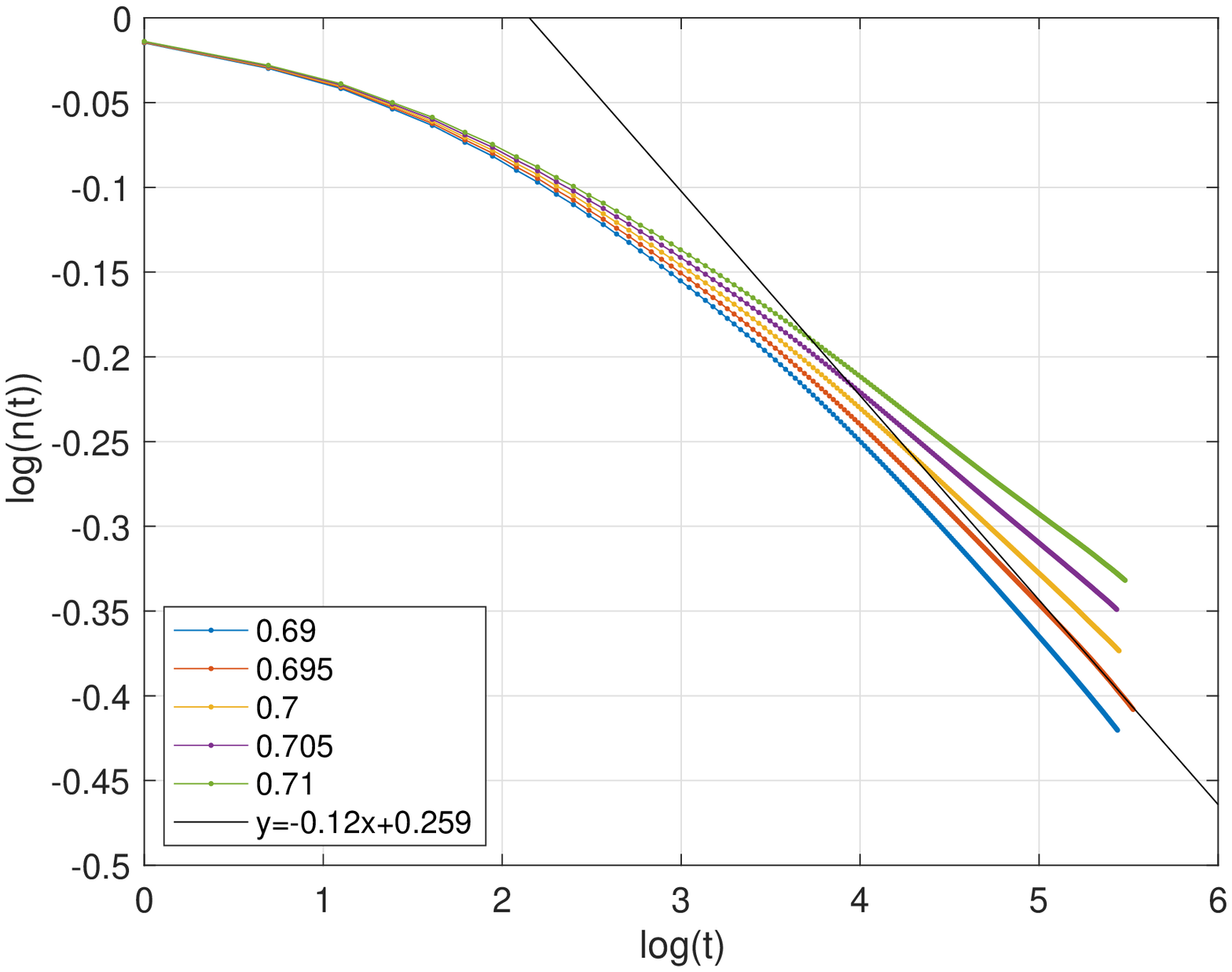}
     \label{fig3cont}}
    \subfloat[][]{
        \includegraphics[scale=0.33]{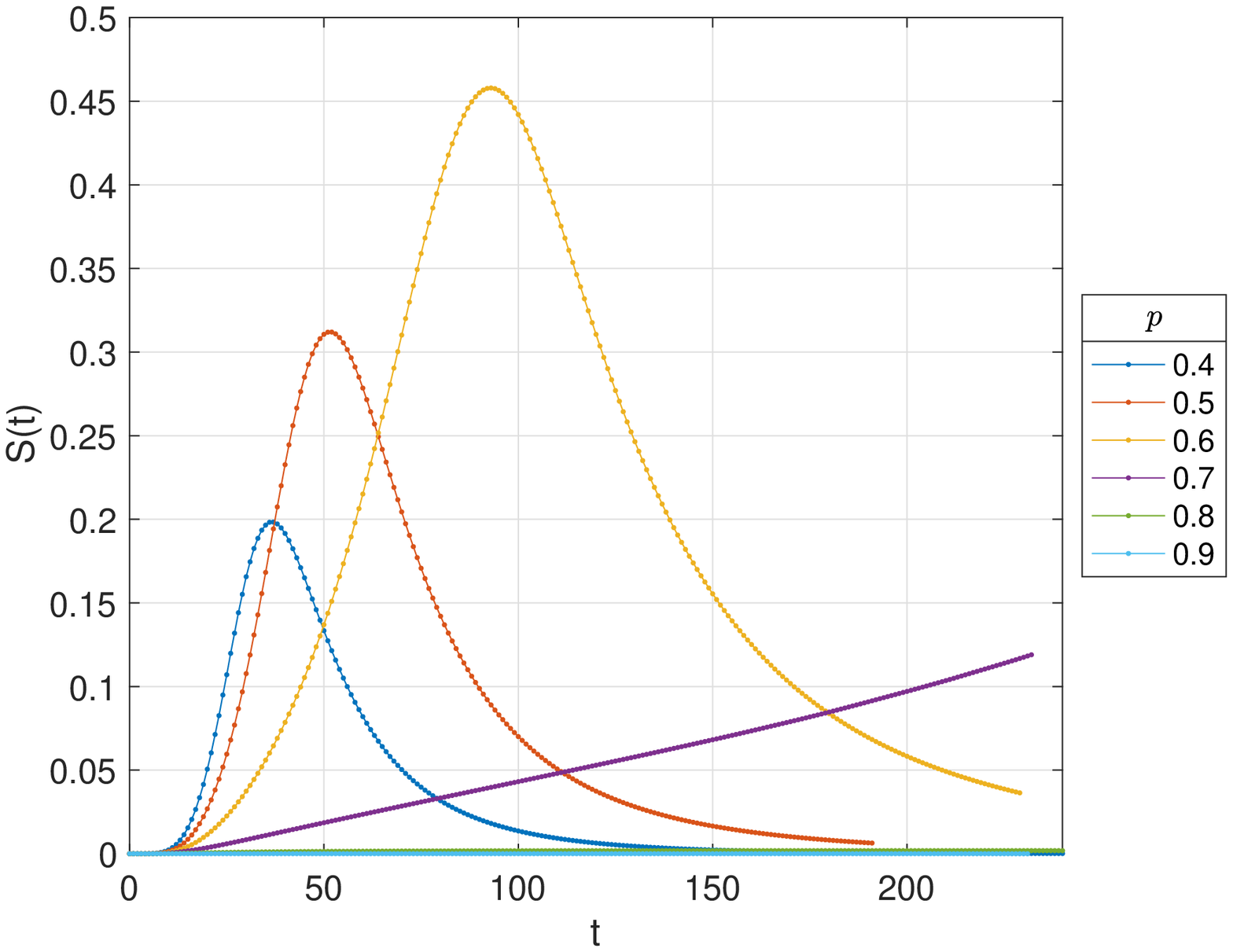}
     \label{fig4cont}}
     \caption{(a) Average number density $n$ (a)-(b) as a function of time $t$ in the small time limit, continuous dynamics limit, $\tau = \sqrt{0.0025/(p(1-p))}$, and no coherent dynamics, $\Omega=0$. (b) The dynamic critical exponent $\delta$ estimated to be $0.12$ by fitting $n(t)\propto e^{-\delta t}$ at the critical point. (c) Entropy $S(t)$ as a function of time. The system was evolved using iTEBD with bond dimension $D=256$ from the fully occupied initial state.}
     \label{fig:ntContOmegaZero}
\end{figure*}

\begin{figure*}
     \centering
     \subfloat[][]{
        \includegraphics[scale=0.33]{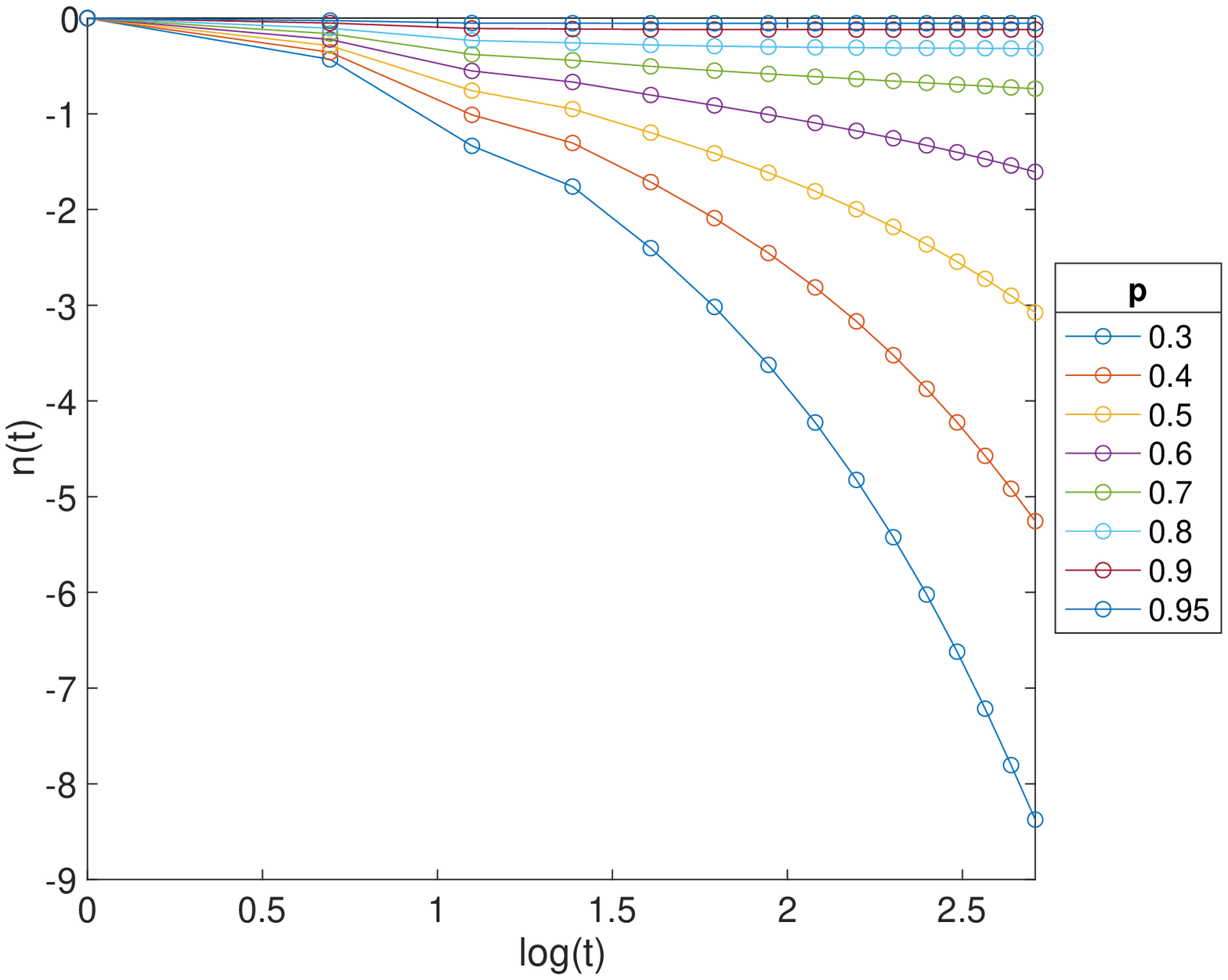}
     \label{fig1disc}}
    \subfloat[][]{
        \includegraphics[scale=0.33]{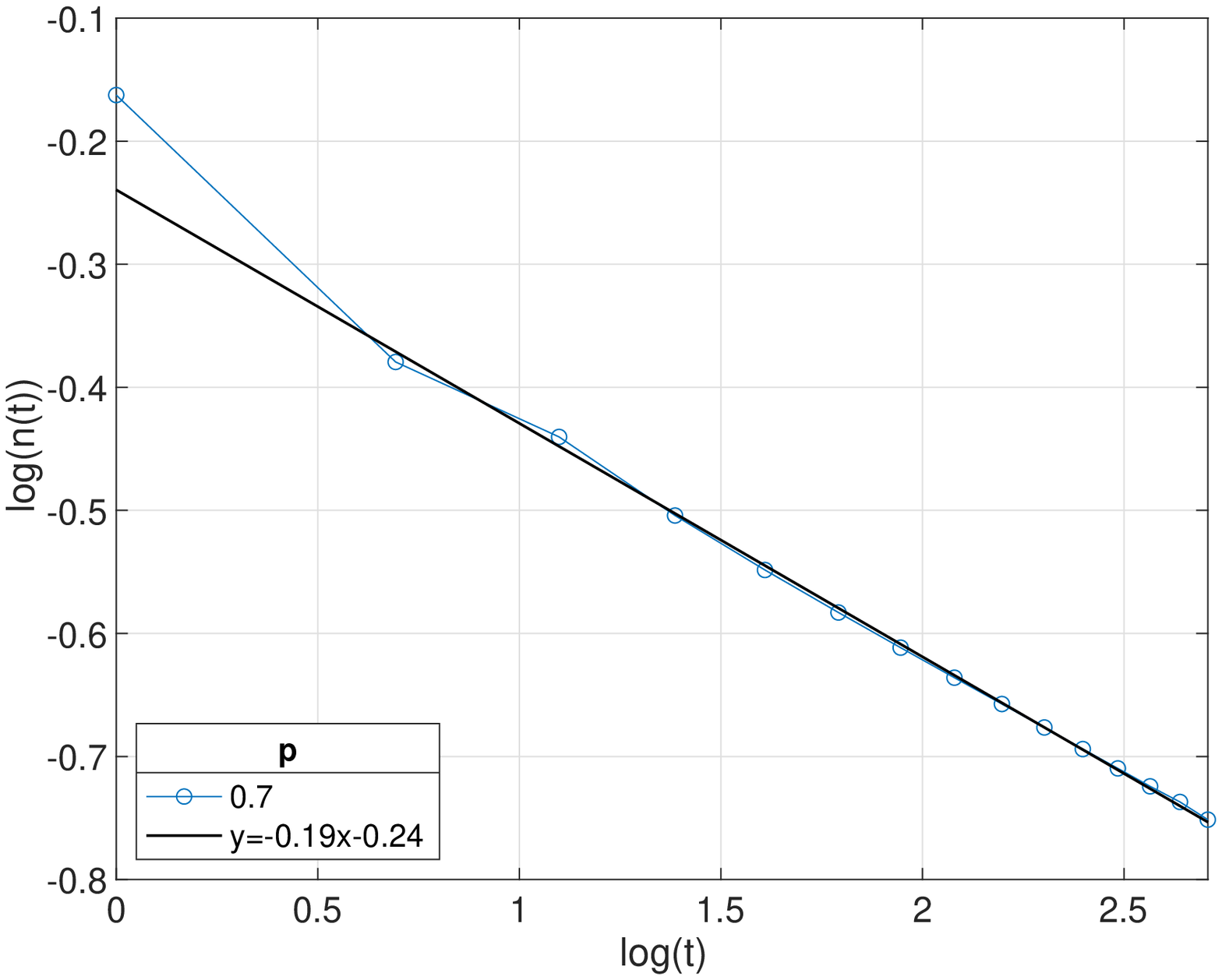}
     \label{fig3disc}}
    \subfloat[][]{
        \includegraphics[scale=0.33]{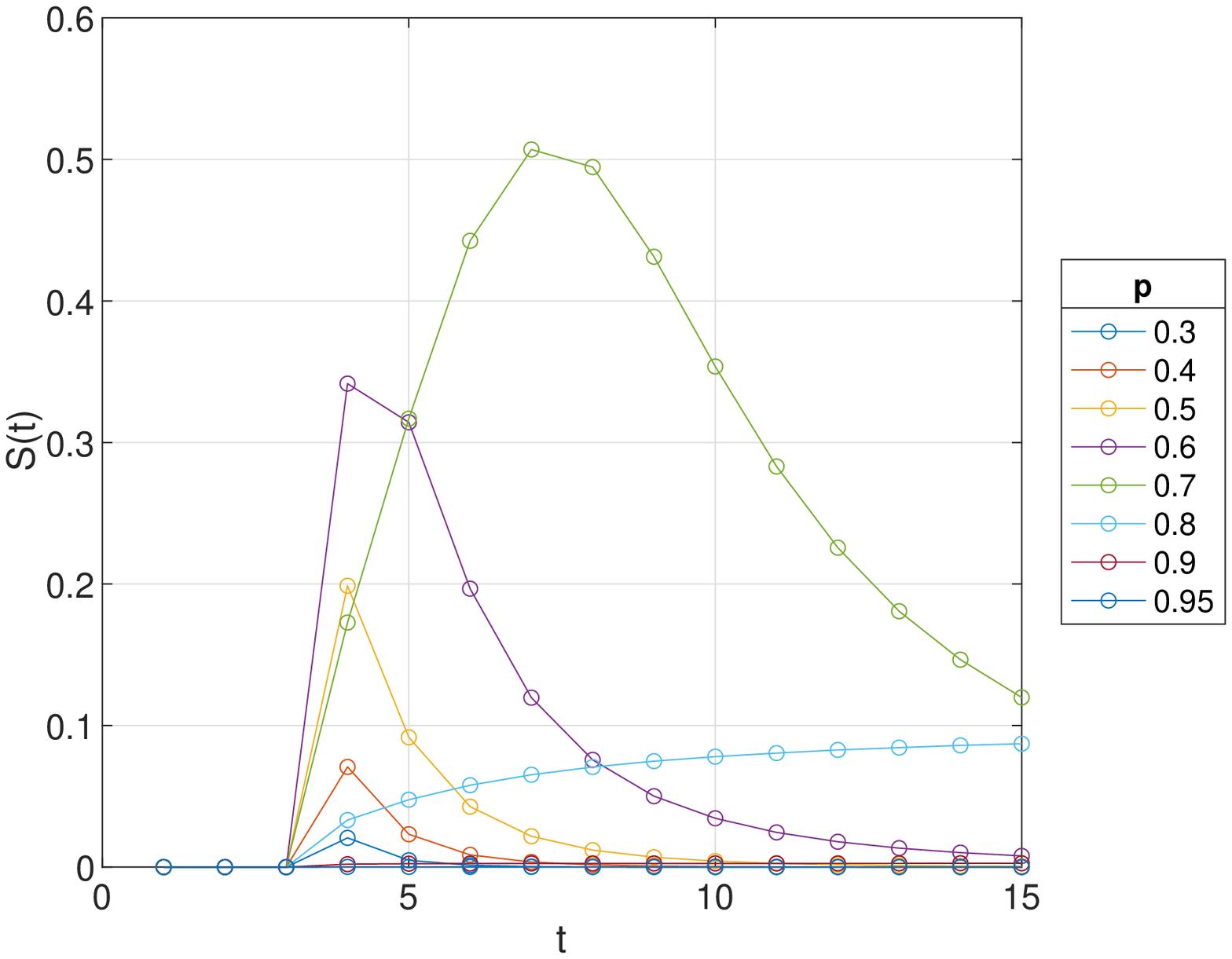}
     \label{fig4disc}}
     \caption{(a) Average number density $n$ (a)-(b) as a function of time $t$ in the continuous dynamics limit, $\tau=10.0$, and no coherent dynamics, $\Omega=0$. (b) The dynamic critical exponent $\delta$ estimated to be $0.19$ by fitting $n(t)\propto e^{-\delta t}$ at the critical point. (c) Entropy $S(t)$ as a function of time. The system was evolved using iTEBD with bond dimension $D=500$ from the fully occupied initial state.}
     \label{fig:ntDiscOmegaZero}
\end{figure*}


From the MPS representing the state of the system at time $t$, we compute two quantities - the average occupation density $n(t)$ and half-chain von Neumann entropy $S$, the evolution of which is well understood for the DP processes \cite{PhysRevLett.123.090601}. The occupation density, given by $n(t) = \text{tr}\left(\hat{n} \rho\right)$ with $\hat{n}=\ket{1}\bra{1}$, can be considered as an order parameter for the transition between absorbing and percolating phases since in the absorbing phase it is zero whereas in percolating phase it is finite. The half-chain entropy is given by $S=(S_{1}+S_{2})/2$ with $S_{1,2} = -\text{tr}[\rho_{1,2}\log \rho_{1,2}]$, where $\rho_1$ and $\rho_2$ are reduced density matrices of half-chain obtained by partitioning at A-B and B-A MPS bonds, respectively. The entropy is computed by first putting the MPS into the canonical form and then using equations $S_{1}=-\sum_j^D{\lambda'^2_{AB}}_j\log {\lambda'^2_{AB}}_j$ and $S_{2}=-\sum_j^D{\lambda'^2_{BA}}_j\log {\lambda'^2_{BA}}_j$, where ${\lambda'_{AB}}_j={\lambda_{AB}}_j/\sum {\lambda_{AB}}_j^2$ and ${\lambda'_{BA}}_j={\lambda_{BA}}_j/\sum {\lambda_{BA}}_j^2$. The evolution of the half-chain entropy is qualitatively different in the absorbing and percolating regimes. The entropy is zero in the absorbing steady state since the absorbing state is a separable product state $\rho=\ket{0}\bra{0}^{\otimes^\infty}$, while in the percolating phase, the entropy of the steady state is finite reflecting the presence of classical correlations.

Figure \ref{fig:ntContOmegaZero} shows the results of the simulations for the continuous QCA model. The results indicate that there is a qualitative change in evolution at $p\sim 0.7$. In the long-time limit the occupation density tends to zero for $p\lesssim 0.7$ and to a finite value for $p\gtrsim 0.7$ as can be seen in figure  \ref{fig:ntContOmegaZero} (a). The QCA evolution is evaluated for a finite number of time-steps and the simulation does not always reach a steady state in the simulated time, particularly, for values of $p$ close to the critical point where the equilibration time is very long. Nevertheless, the transition between absorbing and percolating phase is still clearly detected by evaluating the sign of the curvature of $n(t)$ on a log-log scale i.e. for large $t$, $d\log n/ d t<0$ in the absorbing phase, $d\log n/ d t>0$ in the percolating phase and $d\log n/ d t=0$ at criticality. In figure \ref{fig:ntContOmegaZero} (c), one can see that in the absorbing phase the entropy $S$ initially increases as the correlations are built up in the system, but then falls off to zero as the system approaches the absorbing steady state. The entropy increases monotonically from zero to a finite value in the percolating phase. 

Many non-equilibrium models, including DP, can be described using the phenomenological scaling theory. At the critical point, macroscopic observables, such as correlation length and occupation density, 
are determined by the power scaling laws with characteristic critical exponents. To determine how close the dynamics of our QCA model is to the DP universality class, we have computed the dynamic critical exponent $\delta$, which characterized the critical slowing down of $n(t)$ i.e. $n(t)\propto t^{-\delta}$ at $p=p_c$. The exponent $\delta$ is computed by first locating the critical point and then doing a linear fit of the $\log n(t)$ vs $t$ curve at long time $t$ where the slope of the curve is linear. The set of $n(t)$ curves in the vicinity of the critical point for continuous dynamics is shown in figure \ref{fig:ntContOmegaZero} (b). From this simulation data, we estimate $p_c = 0.695$ and $\delta=0.12$. This is close but not identical to the known values for the site DP, which are $p^{DP}_c=0.705$ and $\delta^{DP} = 0.16$ \cite{doi:10.1080/00018730050198152}.

The results of the simulations of the dynamics in the discrete QCA limit of large $\tau$ are shown in figure  \ref{fig:ntDiscOmegaZero}. The evolution of the average occupation density in the discrete model has the same dynamical features as the continuous model, namely, $n(t)$ decays to zero for $p<p_c$ and to a finite value for $p>p_c$ with $p_c \sim 0.7$. The half-chain entropy has also a characteristic rising and falling behavior in the absorbing phase, and monotonically rising behavior in the percolating phase as shown in figure \ref{fig:ntDiscOmegaZero} (c). Since the Liouvillian acts for a longer time during each update step, it takes fewer steps to reach steady state and the state update is clearly discrete and discontinuous. We find that a higher bond dimension needs to be used for discrete model iTEBD simulation, in order to achieve the same simulation accuracy in the continuous case. For the discrete model, we find that $p_c = 0.71$ and the $\delta=0.19$, which is close to $p^{DP}_c=0.705$ and $\delta^{DP} = 0.16$.




\subsection{Adding the Hamiltonian}
 \begin{figure*}
     \centering
     \subfloat[][]{
        \includegraphics[scale=0.33]{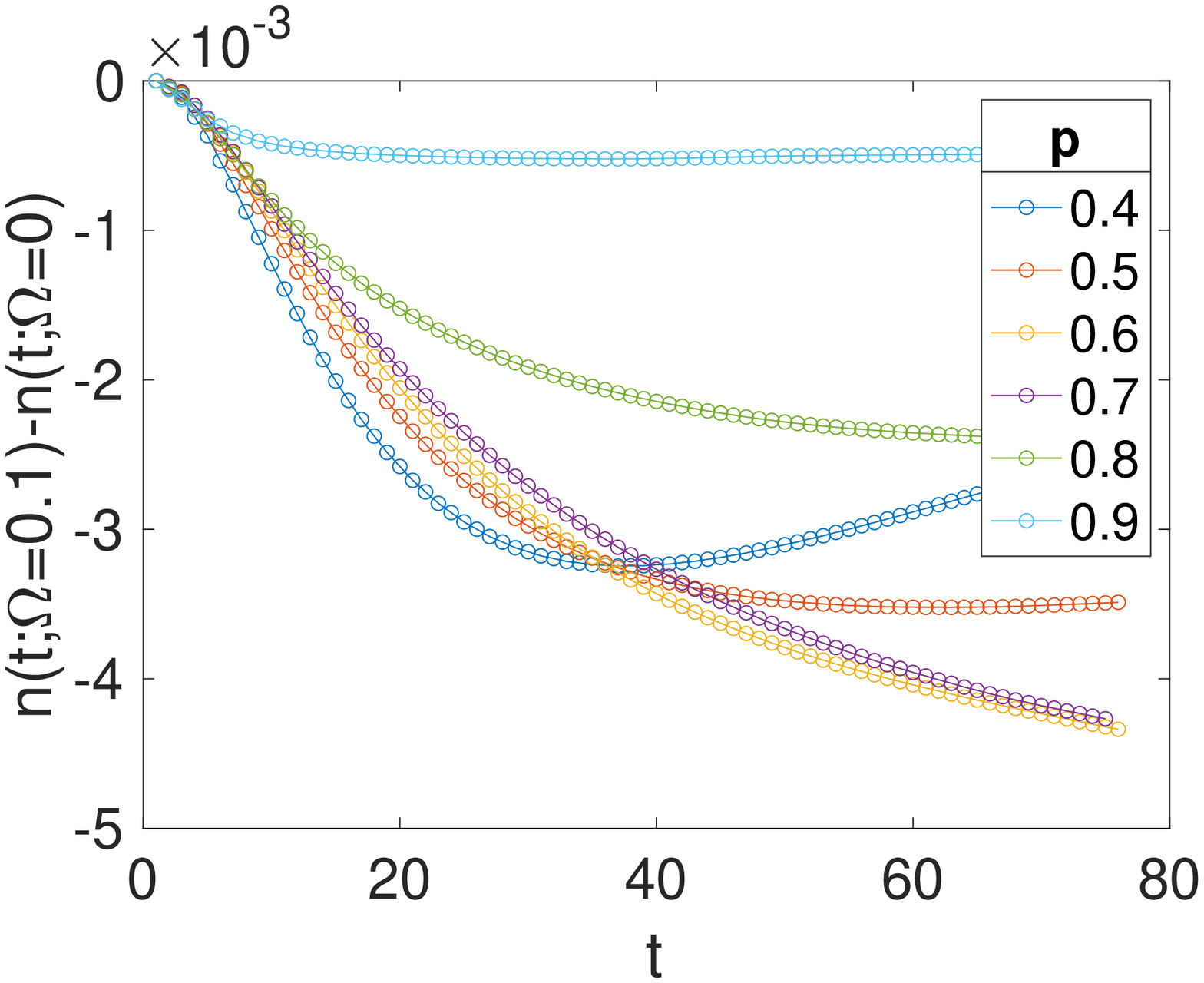}
     \label{fig1disc}}
    \subfloat[][]{
        \includegraphics[scale=0.33]{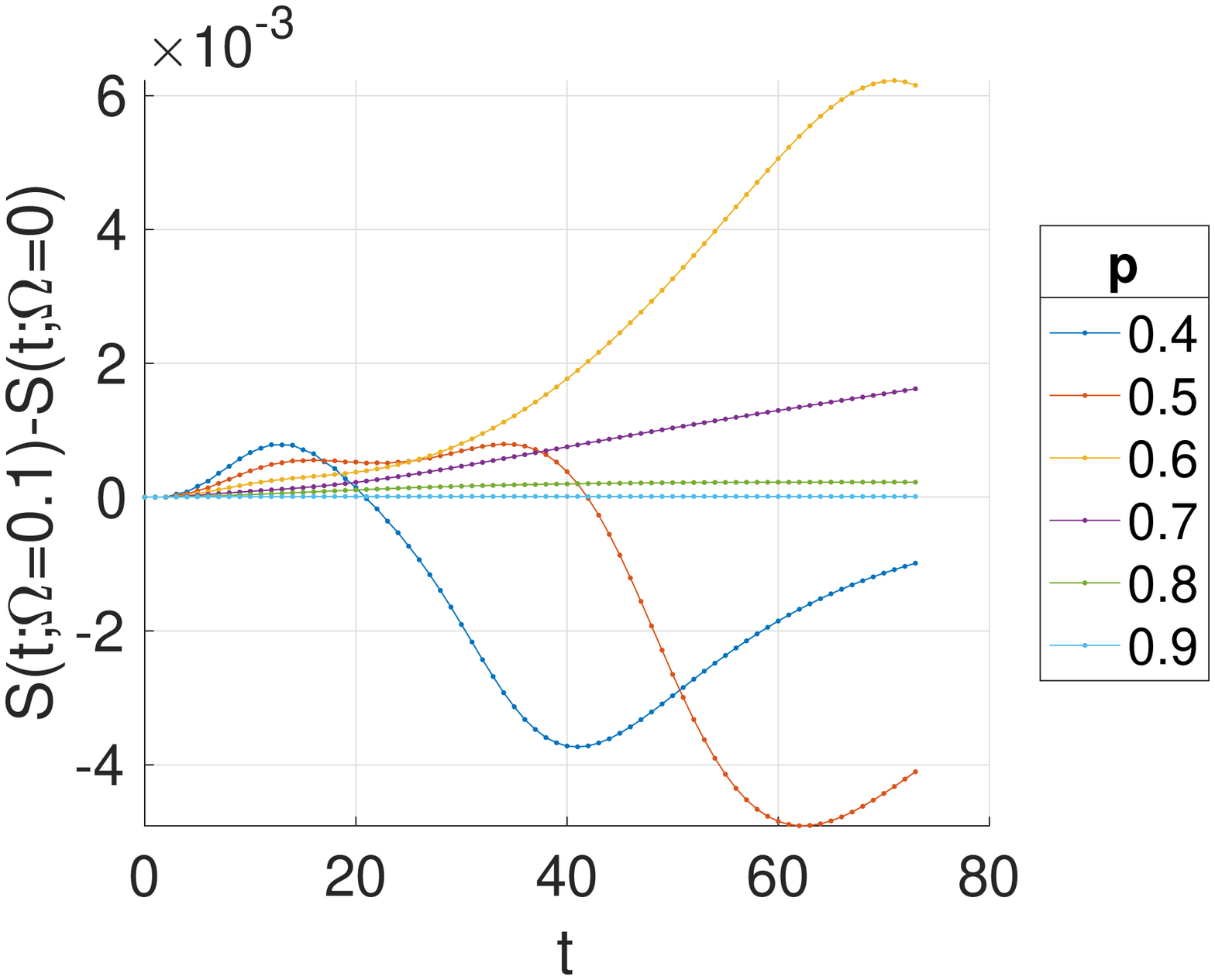}
     \label{fig3disc}}
     \subfloat[][]{
        \includegraphics[scale=0.33]{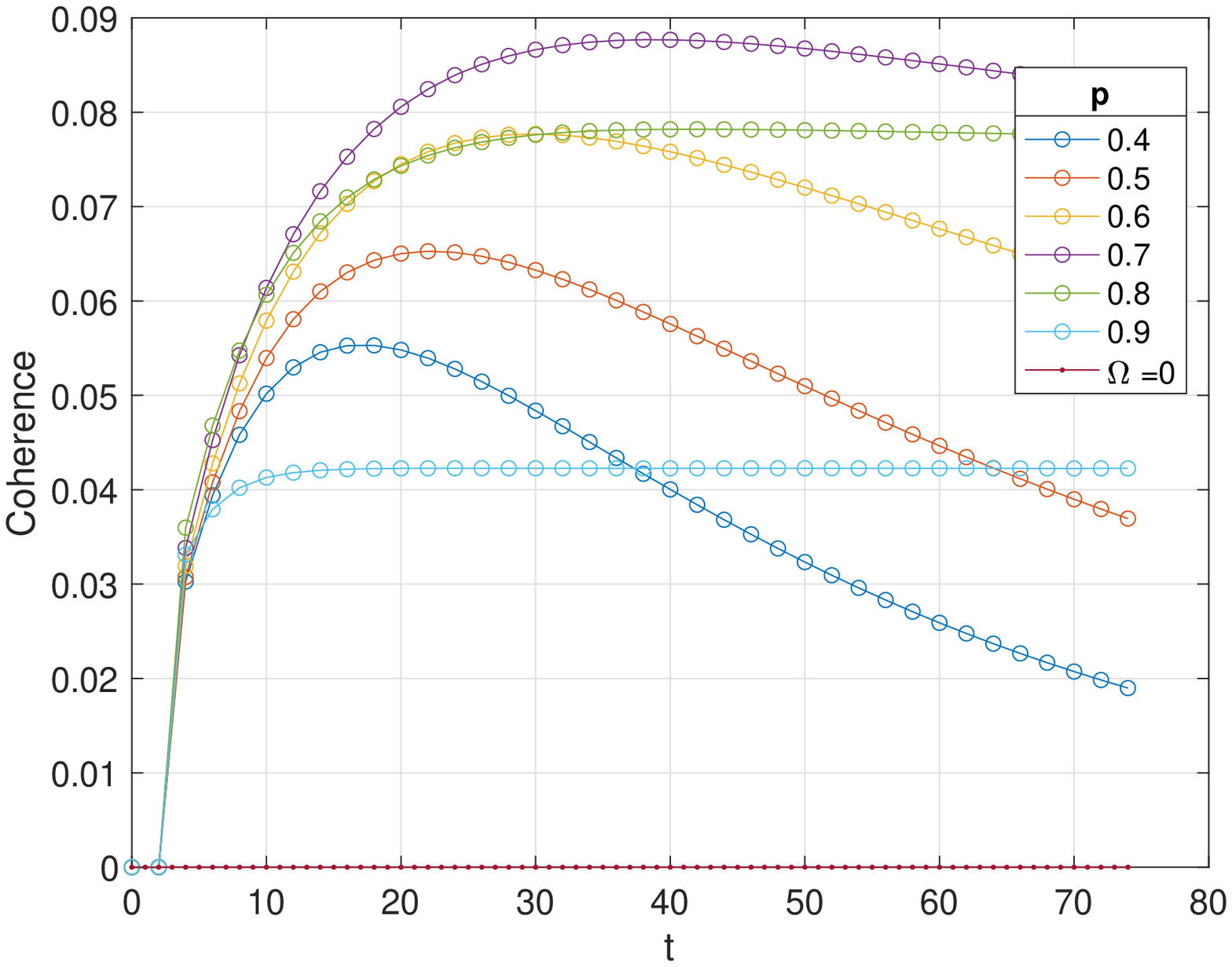}
     \label{fig4disc}}
     \caption{Comparison of the evolution of (a) number density, (b) entropy and (c) single site coherence, $|\rho_{01}^s|+|\rho_{10}^s|$, for different $p$ in the presence ($\Omega=0.1$) and absence ($\Omega=0$) of the Hamiltonian term.  The results are for $\tau = \sqrt{0.0025/(p(1-p))}$; the curves $n(t)$, $S(t)$ for $\Omega=0$ are shown in figure \ref{fig:ntContOmegaZero}.}
     \label{fig:ntContOmega}
\end{figure*}
Having established that our non-unitary QCA model indeed corresponds to directed percolation on a lattice, we now consider adding a local Hamiltonian term to the Liouvillian that generates the dynamical rule. The Hamiltonian can produce coherences and entanglement in the system, and therefore by turning it on, one can systematically explore the influence of quantum effects on the DP non-equilibrium phase transition.  
A Hamiltonian term of the form $\dyad{0} \otimes X \otimes \dyad{0}$ is not compatible with the DK model of DP since the rate $\gamma_{00}$ given by equation (\ref{eq:gammap}) cannot be made positive with $p_{0,0}=0$. 
The local Hamiltonian term, $I \otimes X \otimes I$, is also incompatible with DP since its basis decomposition includes term $\dyad{0} \otimes X \otimes \dyad{0}$. 
For this reason, we will explore the effect of the Hamiltonian, which acts on a central qubit conditioned on the neighbors being in state state $\ket{1}$. The Hamiltonian, $H$, is given by
\begin{equation}
    H = \Omega \dyad{1} \otimes X \otimes \dyad{1}, \label{eq:Ham}
\end{equation}
which corresponds to equation (\ref{eq:Hmodel}) with $\theta_{11}/2=\Omega$ and all other $\theta_{\alpha\beta}$ are set to zero.

The jump rate constants are set to $\gamma_{00}^-=1$, $\gamma_{00}^+=0$, $\gamma_{01}^-=\gamma_{10}^-=1-y$, $\gamma_{01}^+=\gamma_{10}^+=y$, and  $\gamma_{11}^-=1-z$, and
$\gamma_{11}^+=\frac{(2z-1)\gamma_{11}^-+\sqrt{(\gamma_{11}^-)^2-16\Omega^2(1-3z+2z^2)}}{2(1-z)}$. With these rates, the site occupations of the stationary state of the 3-cell rule are identical to the classical DKCA of site DP, where $x=0$, $y=z=p$.

 The results of the iTEBD simulation for the model in the small $\tau$ limit with the Hamiltonian, $\Omega=0.1$, and no Hamiltonian, $\Omega=0$, for various values of $p$ are shown in figure  \ref{fig:ntContOmega}. As can be seen from figure  \ref{fig:ntContOmega} (a) and (b), the evolution of the number densities and the half-chain entropy in the presence of the Hamiltonian are nearly identical, with small differences of the order of $10^{-3}$. 
 The effect of the Hamiltonian dynamics is, however, manifested much more evidently by the presence of coherences in the state of the system.
 As a measure of coherence in the system, we will use the ${{\ell }}_{1}$-norm of coherence, given as ${C}_{1}(\rho )={\min }_{\sigma \in { \mathcal I }}{\parallel \rho -\sigma \parallel }_{{{\ell }}_{1}}={\sum }_{j\ne k}| {\rho }_{j,k}| $, where ${ \mathcal I }$ is the set of all incoherent states and ${\parallel \cdot \parallel }_{{{\ell }}_{1}}$ is the ${{\ell }}_{1}$ matrix norm \cite{Carmeli2018}. Figure \ref{fig:ntContOmega} (c) shows the time evolution of the single site coherence $C_1$ for several values of $p$. As expected, for $\Omega=0$ the coherence is always zero. For $\Omega=0.1$, the coherence is finite during the evolution and its dynamics depends on whether the system is in absorbing, $p<0.7$, or percolating phase, $p>0.7$. For $p<0.7$, the coherence rises initially and then decays towards zero. This is because the Hamiltonian does not act on the absorbing state $\rho=\dyad{0}^{\otimes^\infty}$, and, as the system approaches the absorbing state, the dissipation causes the coherences to gradually decay. For $p>0.7$, the coherence rises and remains finite in a percolating steady state. In the percolating steady state, the rate of generation of coherence and its rate of decay are dynamically balanced. 

Apart from the coherence, an important quantum feature in a many-body dynamical system is entanglement. We have investigated the possibility of entanglement generation by the Hamiltonian given by (\ref{eq:Ham}) by computing concurrence \cite{PhysRevLett.80.2245,PhysRevA.63.044301} in a reduced density matrix of the two adjacent sites, which is a good measure of entanglement for mixed quantum states. We found that the application of a single non-unitary DK QCA 3-cell rule on a product state $\dyad{1}{1}\otimes\dyad{1}{1}\otimes\dyad{1}{1}$ results in a state with non-zero concurrence. However, the concurrence disappears after applying the gates in sequence over the four partitions $\mathcal{P}_1$, $\mathcal{P}_2$, $\mathcal{P}_3$ and $\mathcal{P}_4$, and there is no many-body entanglement present in the steady state in our non-unitary QCA model.

\section{Physical implementation} \label{sec:implementation}

The non-unitary QCA model can be effectively realized experimentally in a one dimensional array of Rydberg atoms or ensembles. In particular, to implement our non-unitary QCA model of DP one can adapt the set-up \cite{PhysRevLett.124.070503}. In \cite{PhysRevLett.124.070503} each site of the one dimensional array is a 3-level system consisting of a ground state $\ket{g}$, strongly interacting Rydberg state $\ket{r}$ and an excited state $\ket{e}$. The sites are equidistant and the interactions are restricted to nearest neighbors. Two laser fields couple the $\ket{g} \leftrightarrow \ket{r}$ and the  $\ket{r}\leftrightarrow\ket{e}$ transitions. The state $\ket{g}$ and $\ket{r}$ will correspond, respectively, to the empty and active states of the DP model. The energy of the Rydberg state depends on the state of the nearest neighbors and the energy level $\ket{r}$ is split into three sublevels $\ket{r_{00}}$, $\ket{r_{10}}$ and $\ket{r_{11}}$, corresponding to the state when the neighboring sites are in states $\ket{g}\ket{g}$, $\ket{r}\ket{g}$ and $\ket{r}\ket{r}$. A multifrequency laser field couples each of the $\ket{g}\leftrightarrow\ket{r}$ and $\ket{e}\leftrightarrow\ket{r}$ transitions with tunable coupling strengths. 
The short-lived $\ket{e}$ decays to $\ket{g}$ with rate $\Gamma$ and is used to implement Lindblad jump operators, which depend on the state of the neighbors. After adiabatically eliminating state $\ket{e}$ the effective master equation, 
is defined by the Hamiltonian of the form of equation (\ref{eq:Hmodel}) and the Lindblad operators are of the form of equation  (\ref{eq:Lmodel}), except there are no $L^+$ jump operator as there are no spontaneous excitation processes in the system.
%
The missing jump operators,    $L_j^{+} = \sum_{\alpha\beta} \sqrt{\gamma_{\alpha\beta}^+}\dyad{\alpha}_{j-1} \otimes \sigma_j^{+}\otimes\dyad{\beta}_{j+1}$, are needed to implement the incoherent excitation of the site $j$ conditional on the states of the neighbors. 
Since $\sigma^+ = X\sigma^-X$, the $L^+$ jump operators can be obtained by conjugating the system with the spin flip operators and then adjusting $\gamma^{\pm}_{\alpha\beta}$ such that they correspond to the desired jump rates. If the time between the application of the flip pulses is small, then the action of $L^+$ and $L^-$ operators in the master equation can be considered to be simultaneous by the Trotterization argument. 

A Rydberg atom implementation of the QCA will necessary involve a finite number of sites. In finite size systems there is no true absorbing/percolating phase transition as the system will always evolve towards an absorbing state as $t\rightarrow\infty$. Nevertheless, one can still see the characteristics of the directed percolation on the timescales which are long compared to the duration of a single QCA step.



\section{Conclusions}
In this paper, we have constructed a one dimensional lattice model with local unitary and non-unitary interactions, which generalizes the classical non-equilibrium process of directed percolation into the quantum domain. 
The model can be viewed as a non-unitary Quantum Cellular Automaton, with a tunable relative strength of the coherent and incoherent dissipative dynamics. 
The 3-cell rule of such non-unitary QCA is the CP map generated by the open non-unitary quantum dynamics specified by the choice of the local Lindblad jump operators and the Hamiltonian. 
Varying the duration $\tau$ of the non-unitary evolution at each time step of the QCA changes the 3-cell CP rule. We have considered the dynamics of the model in the large $\tau$ limit, which corresponds to discrete block partitioned non-unitary QCA, and the small $\tau$ limit, which corresponds to continuous non-unitary QCA. The dynamics of the model in the two limits with no Hamiltonian term was simulated using MPS iTEBD method, and it was found that in both cases the system can undergo absorbing/percolating non-equilibrium phase transitions. The transition point was found to be at $p=0.71$ and $p=0.695$ for the large and small $\tau$ limits, respectively, which is within $0.7\,\%$ of the known critical value of $p$ for the site directed percolation in (1+1)-dimensions. The dynamical exponents $\delta$ were found to be $\sim0.19$ and $\sim0.12$, which appears to be significantly different to the known critical exponent for site DP $\delta=0.16$. However, the discrepancy could be due to the numerical error which increases with time in the iTEBD simulations. Additional simulations using different algorithms would be needed to obtain a more precise value of the critical point and the exponents. 

When adding a local 3-site Hamiltonian term, the dissipation rates can be adjusted such that the evolution of the local site populations is indistinguishable from the purely dissipative dynamics. Nevertheless, such dynamics generates quantum coherences in the system as witnessed by the non-zero off-diagonal elements of the density matrix.
No entanglement was generated by the chosen Hamiltonian and, thus, an interesting open question would to determine whether certain choices of Hamiltonians and Lindblad jump operators would generate a non-unitary QCA rules that produce entangled steady states.

The proposed non-unitary QCA models can be engineered in ultracold atom experiments. We have suggested a possible way of designing the rules in the Rydberg atomic array, based on the proposal in \cite{PhysRevLett.124.070503}. The model in the limit of small $\tau$ is equivalent to the continuous non-unitary QCA and, thus, can be implemented by acting on all sites with identical laser fields. This global control features makes the scheme attractive for experimental implementation.

Since Rydberg arrays can be configured into two dimensional lattice configurations \cite{Browaeys2020}, a natural extension of the present work is to consider (2+1)-dimensional non-unitary QCA. While the generalization of the construction of the non-unitary QCA rules from one to two dimensions is straightforward, the dynamics of the two dimensional CA is usually significantly more complex \cite{Rujan1987}. 
The quantum mechanical generalization of two dimensional CA models is challenging to simulate on classical computers and for this reason their exploration are likely to be driven by experiments. Such models would constitute an excellent test bed for existing and novel tensor network algorithms.

\section{Acknowledgements}
This work was supported by the Australian Research Council Centre of Excellence for Engineered Quantum Systems (Grant No. CE 170100009). This research was undertaken with the assistance of resources from the National Computational Infrastructure (NCI), which is supported by the Australian Government.

\newpage

\newpage
\onecolumngrid
\appendix

\section{Derivation of rate constants $\gamma_{\alpha \beta}^\pm$}	\label{sec:app} 

We present a derivation of the continuous-time version of the DKCA dynamics using the Lindblad formalism.
The principal idea consists of calculating the stationary states after each time step, and setting them in such a way that they obey the dynamics of the desired discrete system.
More specifically, the neighborhood-dependent amplitudes of the Hamiltonian ($\theta_{\alpha\beta}$) and jump operators ($\gamma_{\alpha\beta}^\pm$) are derived in accordance with the update probabilities ($p_{\alpha\beta}$) from the DKCA.

The stationary states are determined by
\ba
    \mathbb L \ket{\rho^s} = 0,
    \label{eq:Lss}
\ea
with $\mathbb L$ being the vectorized version of the Liouvillian in equations \eqref{eq:vectorized} to
\eqref{eq:Dvec} in the main text; writing
\ba
    \mathbb{L}
    &=
        \sum_{\alpha\beta\alpha'\beta'}
        \dyad{\alpha}_{j-1}
        \otimes
        \dyad{\alpha'}_{j'-1}
        \nn\\&\qquad\qquad
        \otimes
        \Big[
            -i\frac{\theta_{\alpha\beta}}{2}
            \l(
                \dyad{0}{1}_j\otimes\dyad{0}_{j'}
                +\dyad{0}{1}_j\otimes\dyad{1}_{j'}
                +\dyad{1}{0}_j\otimes\dyad{0}_{j'}
                +\dyad{1}{0}_j\otimes\dyad{1}_{j'}
            \r)
            \nn\\&\qquad\qquad\qquad
            +i\frac{\theta_{\alpha'\beta'}}{2}
            \l(
                \dyad{0}_j\otimes\dyad{0}{1}_{j'}
                +\dyad{1}_j\otimes\dyad{0}{1}_{j'}
                +\dyad{0}_j\otimes\dyad{1}{0}_{j'}
                +\dyad{1}_j\otimes\dyad{1}{0}_{j'}
            \r)
            \nn\\&\qquad\qquad\qquad
            +\sqrt{\gamma_{\alpha\beta}^+\gamma_{\alpha'\beta'}^+}
            \dyad{0}{1}_j\otimes\dyad{0}{1}_{j'}
            +\sqrt{\gamma_{\alpha\beta}^-\gamma_{\alpha'\beta'}^-}
            \dyad{1}{0}_j\otimes\dyad{1}{0}_{j'}
            \nn\\&\qquad\qquad\qquad
            -\frac{\gamma_{\alpha\beta}^-+\gamma_{\alpha'\beta'}^-}{2}
            \dyad{0}_j\otimes\dyad{0}_{j'}
            -\frac{\gamma_{\alpha\beta}^-+\gamma_{\alpha'\beta'}^+}{2}
            \dyad{0}_j\otimes\dyad{1}_{j'}
            \nn\\&\qquad\qquad\qquad
            -\frac{\gamma_{\alpha\beta}^++\gamma_{\alpha'\beta'}^-}{2}
            \dyad{1}_j\otimes\dyad{0}_{j'}
            -\frac{\gamma_{\alpha\beta}^++\gamma_{\alpha'\beta'}^+}{2}
            \dyad{1}_j\otimes\dyad{1}_{j'}
        \Big]
        \nn\\&\qquad\qquad
        \otimes
        \dyad{\beta}_{j-1}
        \otimes
        \dyad{\beta'}_{j'-1}
\ea
by considering the Hamiltonian \eqref{eq:Hmodel} and jump operators \eqref{eq:Lmodel}
$\forall\alpha,\alpha',\beta,\beta'\in\{0,1\}$, where the indices $j'$ denote the position in the dual space arising from the vectorization approach.
The steady states of the system are then defined by the system of equations
\ba
    \tx{\rom{1}}.
    &\qquad
    0=\frac{-i}{2} (
        \theta_{\alpha\beta}\ \rho_{\alpha,\alpha',1,0,\beta,\beta'}^s
        -\theta_{\alpha'\beta'}\ \rho_{\alpha,\alpha',0,1,\beta,\beta'}^s
        )
    - \frac{\gamma_{\alpha\beta}^- + \gamma_{\alpha'\beta'}^-}{2}\
        \rho_{\alpha,\alpha',0,0,\beta,\beta'}^s
    + \sqrt{\gamma_{\alpha\beta}^+ \gamma_{\alpha'\beta'}^+}\
        \rho_{\alpha,\alpha',1,1,\beta,\beta'}^s
    \nn\\
    \tx{\rom{2}}.
    &\qquad
    0=\frac{-i}{2} (
        \theta_{\alpha\beta}\ \rho_{\alpha,\alpha',1,1,\beta,\beta'}^s
        -\theta_{\alpha'\beta'}\ \rho_{\alpha,\alpha',0,0,\beta,\beta'}^s
        )
    - \frac{\gamma_{\alpha\beta}^- + \gamma_{\alpha'\beta'}^+}{2}\
        \rho_{\alpha,\alpha',0,1,\beta,\beta'}\textbf{}
    \nn\\
    \tx{\rom{3}}.
    &\qquad
    0=\frac{-i}{2} (
        \theta_{\alpha\beta}\ \rho_{\alpha,\alpha',0,0,\beta,\beta'}
        -\theta_{\alpha'\beta'}\ \rho_{\alpha,\alpha',1,1,\beta,\beta'}
        )
    - \frac{\gamma_{\alpha\beta}^+ + \gamma_{\alpha'\beta'}^-}{2}\
        \rho_{\alpha,\alpha',1,0,\beta,\beta'}^s
    \nn\\
    \tx{\rom{3}}.
    &\qquad
    0=\frac{-i}{2} (
        \theta_{\alpha\beta}\ \rho_{\alpha,\alpha',0,1,\beta,\beta'}^s
        -\theta_{\alpha'\beta'}\ \rho_{\alpha,\alpha',1,0,\beta,\beta'}^s
        )
    - \frac{\gamma_{\alpha\beta}^+ + \gamma_{\alpha'\beta'}^+}{2}\
        \rho_{\alpha,\alpha',1,1,\beta,\beta'}^s
    + \sqrt{\gamma_{\alpha\beta}^- \gamma_{\alpha'\beta'}^-}\
        \rho_{\alpha,\alpha',0,0,\beta,\beta'}^s,
    \label{eq:systemofequations3}
\ea
where the first and last two subscripts of the steady state elements $\rho_{\alpha,\alpha',c,c',\beta,\beta'}^s$ 
denote the basis of the left or right sites, $(j-1,j'-1)$ or $(j+1,j'+1)$ respectively,
while the center site $(j,j')$ is associated with $c,c'\in\{0,1\}$.

W.l.o.g.~one can consider that the nearest-neighbor states and their dual space correspondence are identical, setting $\alpha=\alpha'$ and $\beta=\beta'$.
The Liouvillian is then simplified to 
\ba
    \mathbb{L}
    &=
        \sum_{\alpha\beta}
        \dyad{\alpha}_{j-1}
        \otimes
        \dyad{\alpha}_{j'-1}
        \nn\\&\qquad\qquad
        \otimes
        \Big[
            \frac{-i\theta_{\alpha\beta}}{2}
            \Big(
                \dyad{0}{1}_j\otimes\dyad{0}_{j'}
                +\dyad{0}{1}_j\otimes\dyad{1}_{j'}
                +\dyad{1}{0}_j\otimes\dyad{0}_{j'}
                +\dyad{1}{0}_j\otimes\dyad{1}_{j'}
            \nn\\&\qquad\qquad\qquad\qquad\qquad
                -\dyad{0}_j\otimes\dyad{0}{1}_{j'}
                -\dyad{1}_j\otimes\dyad{0}{1}_{j'}
                -\dyad{0}_j\otimes\dyad{1}{0}_{j'}
                -\dyad{1}_j\otimes\dyad{1}{0}_{j'}
            \Big)
            \nn\\&\qquad\qquad\qquad
            +\gamma_{\alpha\beta}^+ \,
            (
            \dyad{0}{1}_j\otimes\dyad{0}{1}_{j'}
            -
            \dyad{1}_j\otimes\dyad{1}_{j'}
            )
            \nn\\&\qquad\qquad\qquad
            +\gamma_{\alpha\beta}^- \,
            (
            \dyad{1}{0}_j\otimes\dyad{1}{0}_{j'}
            -
            \dyad{0}_j\otimes\dyad{0}_{j'}
            )
            \nn\\&\qquad\qquad\qquad
            -\frac{\gamma_{\alpha\beta}^-+\gamma_{\alpha\beta}^+}{2} \,
            (
            \dyad{0}_j\otimes\dyad{1}_{j'}
            +
            \dyad{1}_j\otimes\dyad{0}_{j'})
        \Big]
        \nn\\&\qquad\qquad
        \otimes
        \dyad{\beta}_{j-1}
        \otimes
        \dyad{\beta}_{j'-1}.
\ea
For clarity the subscripts $\alpha,\beta$ are omitted further by replacing
$\gamma_{\alpha\beta}^\pm \rightarrow \gamma^\pm$,
$p_{\alpha\beta} \rightarrow p$,
$\theta_{\alpha\beta}/2 \rightarrow \Omega$,
and
$\rho_{\alpha,\alpha',c,c',\beta,\beta'}^s
=\rho_{\alpha,\alpha,c,c',\beta,\beta}^s
\rightarrow \rho_{c,c'}^s$;
yielding a simplified system of equations that define the steady state elements $\forall\alpha,\beta$:
\ba
    \tx{\rom{1}}.&\qquad 0=i\Omega(\rho_{01}^s-\rho_{10}^s) - \gamma^-\rho_{00}^s+\gamma^+\rho_{11}^s    \nn\\
    \tx{\rom{2}}.&\qquad 0=i\Omega(\rho_{00}^s-\rho_{11}^s) - \frac{\gamma^++\gamma^-}{2}\rho_{01}^s      \nn\\
    \tx{\rom{3}}.&\qquad 0=i\Omega(\rho_{11}^s-\rho_{00}^s) - \frac{\gamma^++\gamma^-}{2}\rho_{10}^s      \nn\\
    \tx{\rom{4}}.&\qquad 0=i\Omega(\rho_{10}^s-\rho_{01}^s) + \gamma^-\rho_{00}^s-\gamma^+\rho_{11}^s,
    \label{eq:systemofequations}
\ea
where it is to note that equations
\rom{2.} and \rom{3.}
demand $\rho_{01}^s=(\rho_{10}^s)^*$,
and
\rom{1.} and \rom{4.}
lead to $\rho_{00}^s+\rho_{11}^s=1$ representing the trace-preserving property of the reduced density matrix at site $j$.

Desired dynamics are thus encoded into the parameters that describe the system, $\gamma^\pm$ and $\Omega$.
As $\Omega$ describes the strength of the Hamiltonian which tunes quantum coherences, this shall not be fixed, but remain a variable to effectively increase or decrease the quantum coherences, turning them on $(\Omega\neq0)$ or off $(\Omega=0)$.

Here, the stationary states of the system are set after each time step to obey the dynamics of the DKCA.
In this model an arbitrary input state is taken to 
$\begin{pmatrix}
        p&0\\
        0&1-p
\end{pmatrix}$,
i.e.~the system is in the classical `1' state with probability $p$, or in `0' with probability $1-p$ after one time step.
The corresponding quantum dynamics are described in the Lindblad formalism by setting $\rho_{11}^s=1-p$.
and then solving the system of equations \eqref{eq:systemofequations} for one of the variables that describe the system, say $\gamma^+$.
By introducing the Hamiltonian part of the Liouvillian quantum coherences show up as the off-diagonal matrix elements of the density matrix:
    $\begin{pmatrix}
        p & \rho_{01}^s
        \\
        -\rho_{01}^s& 1-p 
    \end{pmatrix}$
with
\ba
    \rho_{01}^s
     &= \begin{cases}
         0
         &\tx{, if }\Omega=0
         \\
        \frac{-i}{4\Omega}\l(\gamma^- 
        - \sqrt{(\gamma^-)^2-16\,\Omega^2\,(1-3p+2p^2)}\r)
        &\tx{, if }\Omega\neq0
        \end{cases},
\ea
where
\ba
    \gamma^+
    &= \begin{cases}
        \frac{p}{1-p}\gamma^-
         &\tx{, if }\Omega=0
         \\
        \frac{(2p-1)\gamma^-+\sqrt{(\gamma^-)^2 - 16\,\Omega^2 (1-3p+2p^2) }}{2(1-p)} &\tx{, if }\Omega\neq0
        \end{cases},
\ea
which is e.g.~with $\gamma^-=1$
only real and positive in the cases
\ba
    \tx{\rom{1}}.&\qquad \Omega=0   \nn\\
    \tx{\rom{2}}.&\qquad p\geq\frac{1}{2} \ \forall\,\Omega  \nn\\
    \tx{\rom{3}}.&\qquad p<\frac{1}{2} \tx{ if } 
            \Omega  \leq \frac{1}{16(1-3p+2p^2)}.
\ea

Note that for our simulations of the DP model the Hamiltonian is chosen to be only turned on if both neighboring states are in the `1'-state, i.e.~we set $\Omega=\theta_{11}$/2 where $\alpha=\beta=1$.

\end{document}